\documentstyle[aps,epsfig]{revtex}                         

\draft
\begin{document}
\twocolumn[\hsize\textwidth\columnwidth\hsize
     \csname @twocolumnfalse\endcsname



\title{Information transfer via the phase: \\
       A local model of Einstein-Podolsky-Rosen experiments.
       }

\author{W. A. Hofer}
\address{
         Dept. of Physics and Astronomy, University College London \\
         Gower Street, London WC1E 6BT, E-mail: w.hofer@ucl.ac.uk}

\maketitle

\begin{abstract}
Conventionally, one interprets the correlations observed in
Einstein-Podolsky-Rosen experiments by Bell's inequalities
and quantum nonlocality.
We show, in this paper, that identical correlations arise, if the phase
relations of electromagnetic fields are considered.
Conceptually, we proceed as follows: First, it is proved that
non-factorizability does not mean nonlocality. This is done
by a one photon model. Then, it is shown, that the ''classical''
expression for the correlation sums up photons of different pairs
indiscriminately. This feature accounts for the lower correlations in
the ''classical'' model. And finally, an integral is derived, including
the properties of both photons while retaining the linearity of
fields between the polarizers. This expression describes the
measured values correctly. It seems thus that
quantum nonlocality can be understood as a combination of linearity
of possible electromagnetic fields between the polarizers
and a relation of the electromagnetic fields of the two photons
via a phase. We expect the same feature to arise in every experiment,
where joint probabilities of separate polarization measurements are
determined.
\end{abstract}


\vskip2pc]

The existence of correlations between particles in space-like separation
is arguably the single most important problem in quantum optics. The
literature about these Einstein-Podolsky-Rosen (EPR) experiments and
their practical realization is impressive \cite{epr35,afriat99,bell64}.
Why is this so? Since Aspect's first measurements \cite{aspect82}
experimenters have sought to eliminate all possible loopholes restricting
the generality of the results. In the latest set of measurements, performed
under strict Einstein locality conditions
\cite{weihs98}, a violation of Bell's inequalities  \cite{bell64} was observed,
although the two measurement devices
were separated by 400 m. In common terms: the measurement of a property of
photon 1 is correlated to the measurement of the same property of photon 2,
although the two particles have no possibility to influence each other by any
type of conventional field. In his recent review Aspect concludes from the
experimental facts that \cite{aspect99} " ... it is impossible to assign local
physical reality to each photon." There is no way, how experimental results
can be reconciled with the image of two single photons propagating in
opposite directions, where they undergo separate measurements. Consequently,
some work has been devoted recently to a careful analysis of the Bell inequalities
and an assessment of their validity or invalidity \cite{peres99,sica99}.
But even if the Bell inequalities are flawed, the experimental facts
remain the same: correlations exist, and we don't know, why.

\begin{figure}
\begin{center}
\epsfxsize=1.0\hsize
\epsfbox{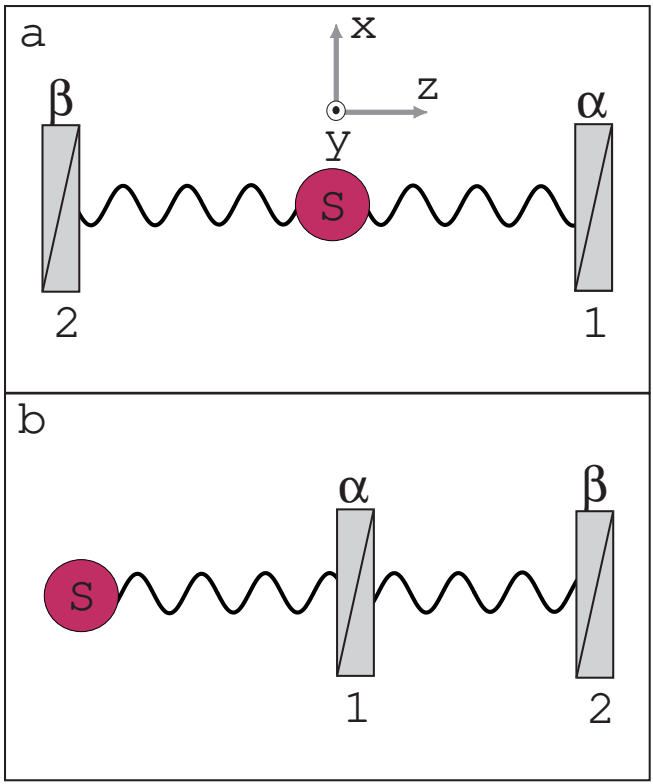}
\end{center}
\vspace{0.5 cm}
\caption{
         One dimensional model of EPR type experiments. {\bf a}
         The measuring devices 1 and 2 are in opposite directions
         from the photon source S. The polarizers are set to the angles
         $\alpha$ and $\beta$, respectively.  {\bf b} Successive polarization
         measurement on a single photon emitted from S.
        }
\label{fig001}
\end{figure}

In essence, quantum nonlocality derives from the conditional probabilities of
two polarization measurements with the polarizer angles set to $\alpha$ and
$\beta$, respectively (see Fig. \ref{fig001} {\bf a}). The probability of a
coincidence at both devices is described by a term  sin$^2$($\beta - \alpha$)
\cite{weihs98}. Formally this term  represents  the expectation value of
spin-measurements in quantum mechanics. However, there exists no known field
which allows for a physical connection between the two measuring devices.
For this reason the question of information transfer is still unsolved.
We propose, in this report, a solution based on the phase of the photon's
electromagnetic field. And the question, how the two measurements can be
related without any physical medium, will be answered approximately by:
the photons remember their initial phase at their common origin. It is
this memory, which shows up in the ticks of the detectors and coincidence
rates.

The measurements are performed using laser pulses, ideally of very short duration,
which are converted in an optical converter into two separate pulses of
lower frequency and a defined phase relation \cite{zeilinger99}. The
coherence length of the pulses is very high \cite{weihs98}. Due to the
short duration of the pulses the electromagnetic fields and their field
characteristics are limited in space. But given the high coherence length,
their phase along the photon's path of propagation remains defined and
in general equal to the phase of a hypothetical electromagnetic field,
covering the whole experimental setup. It is this quality of the laser pulses,
which is physically decisive for correlation measurements. In this sense the
experiments are of an admirable precision and at the limit of the
experimentally feasible today.

In a theoretical treatment reflecting this quality of coherence
(also present, e.g. in the representation in quantum mechanics) the
electromagnetic field of the single photons and the hypothetical
electromagnetic field throughout the system become interchangeable.
Their only difference is a scale for the amplitudes. This difference,
as will be shown presently, is of no effect on obtained results.
In the current treatment we will calculate the field aspects of the
correlation measurements by using this hypothetical field.
To clarify the issues at hand we initially model two successive polarization
measurements on a single photon (see Fig. \ref{fig001} {\bf b}). This model
will then be generalized to account for coincidence measurements of two
photons. The hypothetical field vector of the electric field shall be polarized
in $x$-direction, the amplitude is E$_{0}$. A straightforward calculation of
the electric field after two polarization measurements yields a reduced and
rotated amplitude. The amplitude E$_{2}$ after the measurement with device 2
will be:

\begin{equation}\label{001}
E_{2} = \cos \alpha \cos (\beta - \alpha) \left|
\cos \beta {\bf e}_{x} + \sin \beta {\bf e}_{y} \right| E_{0}
\end{equation}

The transformation T($\alpha,\beta$) with $ E_{2} = T(\alpha,\beta) E_{0}$
describes the events in three discrete steps: (i) A reduction of the field at
the first device (cos $\alpha$). (ii) A conditional reduction of the amplitude
at the second device (cos($\beta - \alpha$)). (iii) A rotation of the
amplitude vector ($\cos \beta {\bf e}_{x} + sin \beta {\bf e}_{y}$).
The decisive step is step (ii): the conditional reduction of the amplitude at
device 2 given the measurement at 1. The term combines two operations: the rotation
of the field vector at device 1, and the reduction of the amplitude at device 2.

It is essential, for an understanding of the following, that this term combines
physical events at two different locations. The physical content of the
mathematical expression can be visualized. Two polarizers with perpendicular
planes of polarization black out all light. But a third, in a diagonal
plane of polarization between the former two, is sufficient to let a
part of the original light pass all three polarizers. Although the effect
seems puzzling to the layman, it is nonetheless completely understandable.
It shows, what could be called the memory of the photon's
electromagnetic field: for the outcome of a polarization measurement
the previous treatment is generally relevant. The conditional probability
equally must be expressed in terms of settings at different locations.

If the measurement is repeated with polarizers set to the angles
$\gamma, \delta, \epsilon$ etc, then
the representation contains terms of the form cos($\gamma - \beta$)
cos($\delta - \gamma$) cos($\epsilon - \delta$) etc. If the probability
of a measurement is proportional to the intensity of the electromagnetic field,
which it commonly is in measurements using cascade detectors, then the reduction
of the probability in two consecutive measurements is equal to
cos$^2$($\beta - \alpha$). But this means, that we can calculate the conditional
probability of the two measurements. The result is:

\begin{equation}
P(\alpha,\beta) = 1 - \cos^2 (\beta - \alpha) =  \sin^2 (\beta - \alpha)
\end{equation}

Since the reduction is a relative measure, the actual numerical value of the
amplitude is not decisive. Therefore the conditional probability calculated
from this (hypothetical) electromagnetic field and the one computed from the
actual photon field are strictly equivalent.
The result is also equal to the one quoted above and which is the standard result
in EPR-type experiments. Although the mathematical expression suggests nonlocality,
the actual physical process can be completely local. It is only required that the photon
has a finite volume \cite{hofer98} and that the two measuring devices are sufficiently
far apart. These conditions are met in most state-of-the-art experiments
\cite{weihs98}. Note that the correlation probability, in this one photon
model, cannot be factorized into events at device 1 and device 2. But that
does not mean an involvement of nonlocality: the photon is perfectly localized
at every given moment along its path.

That this feature is also inherent in two photon model is indicated by the fact
that the one-photon model (determine the probability of measurement 2
{\it after} 1 has been performed) is an exact representation of the probability
calculus of the two photon case
(determine the probability of 2 {\it under the condition that} 1
has been performed). The only changes in a two photon model should therefore
be the initial phase between the two photons and a factor of two for the
detector counts and probabilities.
Even with a space-like separation of the two devices
\cite{weihs98}, the correlation probability then contains a connection
between settings at device 1 and 2.

Unfortunately, there is no way to actually apply the equivalence. The
reason is simple: polarizations with arbitrary angles follow, classically,
a $\cos^2 \phi$ characteristic. In this context it should be noted that we
employ, in the following, the term phase in both of its meanings: it can,
e.g. for circular polarization, be the actual angle of the field vector
at a given moment, or it describes the propagation of the photons along
their path from the source to the polarizers. Since these two variables
are connected by the wave features of the photons, a separation seems
unnecessary. On the other hand, using a common symbol simplifies the
notation considerably.
Returning to single polarizations, it is clear that treating each
measurement separately destroys the vital phase relations. Why are
these relations important? Consider
the two photon model where the electromagnetic fields of the photons
are related by a phase at their origin. Given an experimental setup,
the hypothetical electromagnetic field, accounting for the coherence
of the photon beams, is exactly defined by the boundary conditions at
the two polarizers. This derives from the time inversion symmetry of
the wave equation. Now consider the way, correlation probabilities are
treated e.g. in Furry's model \cite{furry36}. The integral over all
measurements is:

\begin{equation}
P (\alpha,\beta) = \int d x \cos^2(x) \cos^2(x - (\alpha - \beta))
\end{equation}

The integral clearly describes the impacts, but does so by completely separating
the two measurements. There is no way to guarantee, in this expression, that
the electromagnetic fields are indeed linear throughout the space between the
polarizers. To the point: the integral combines not only measurements, which
belong to the same photon pair, but also measurements, where one measurement
belongs to a different pair than the other. And in the latter case the
phase between events is, of course, arbitrary. However, the integration
does not completely destroy the correlations one expects if only
photons of the same pair are considered. The remainder of an interference
pattern is still obtained. It has been said, in this
respect, that quantum correlations are stronger than classical ones.
This seems not quite right. Only in quantum mechanics the correlations
are computed by an expression conserving the linearity and thus guaranteeing,
that photon 1 and photon 2 really belong to the same pair.

A physical approach to this problem, which could be called the
analysis of the EPR problem from a field theoretical and statistical
angle, can be based on a non standard photon model \cite{hofer98}. In this
model transversal properties are described by electromagnetic fields, while
longitudinal properties are related to the kinetic energy of the photons.
To account for both features the fields must be described by complex vectors.
In the same spirit the polarization measurement is performed on a complex
function, where the electromagnetic component is $\cos \phi$ and the kinetic
component $i \sin \phi$. In electrodynamics complex fields are standard
practise, as is the computation of intensity from the real part of the fields alone.
In a two photon measurement, where the fields between polarizers preserve their
full linear features, only possible relations between the two measurements
must be included. Which means, that the square of the fields cannot
be computed for each measurement alone. It has to be calculated including
interference terms. An integral along these lines would be the following:

\begin{eqnarray}
I (\alpha, \beta) = \int_{0}^{2 \pi} dx
\left| \left[\cos(x - \alpha) + i \sin(x - \alpha) \right] \right.\\
     \left.  \left[\sin(x - \beta - \phi_{0}) + i \cos(x - \beta - \phi_{0}) \right]
\right|^2 \nonumber
\end{eqnarray}

The exchange of $\sin$ and $\cos$ in the field of the second photon signals
a phase shift of $\pi/2$. We set $\phi_{0}$ to zero for convenience.
Now if the correlation probability, or the correlated change of
intensity in the measurements, is taken from the real part of this
integral, we arrive for the normalized integral at an expression
already deduced by Kracklauer, and which is equivalent to the measured
correlations \cite{kracklauer99}:

\begin{equation}
\bar{P} (\phi) =
\frac{\int_{0}^{2 \pi} dx \left[\cos x \sin (x - \phi) -
                               \sin x \cos (x - \phi) \right]^2
              }{2 \int_{0}^{2 \pi} dx \left[ \cos^2 x + \sin^2 x \right]
               }
\end{equation}

In the general case this leads to the following expression for the
average correlation probability:

\begin{equation}
\bar{P} (\alpha, \beta) = \frac{1}{2} \sin^2 (\beta - \alpha - \phi_{0})
\end{equation}

Here $\phi_{0}$ denotes the setup and the type of measurement (either
transmission or adsorption as the relevant events).
Can we justify the two operations? (i) Taking only the real part of the
complex valued function. (ii) Linking it to intensity. The manipulations
are strictly valid only in the hypothetical field, because only in this
case the boundary conditions apply. In terms of classical expressions,
where the intensity usually is computed in this way, there seems no problem.
The expression again is a relative one, so it should also apply to the
single photons and their measurements.

Physically speaking, one has to make sure that the integration contains
only photons belonging to the same pair. A way to achieve this, is the
linearity requirement. Its ultimate mathematical foundation is the
linearity of both theories: quantum mechanics and electrodynamics.
In quantum mechanics this is done by describing the measurements locally,
and by retaining the phase between the two points of measurement
\cite{unni00}. A method to the same effect in electrodynamics is
to retain the linearity in the integral, and to compute the square
of the real part. The chosen method seems to be rather a matter of
taste, once the physics of the problem is clear. And the physics
behind both treatments is the same: a phase connection between the
two separate photons due to the high coherence of the beam.

The conceptual steps leading to this result can be rephrased as follows.
First it was proved that non-factorizability does not mean nonlocality.
This was done by the one photon model. Then it was shown, that the
''classical'' formulation sums up photons of different pairs
indiscriminately. And finally, using a non standard model of photons,
we provided a formulation for the integral including the features
of both photons while retaining linearity. The last step can probably
be solved in a different way. And it is, in fact, in quantum
mechanics. This is not decisive. Decisive is the shift of focus.
While previously the main question was:

\begin{itemize}
\item
{\it How can there be a nonlocal connection between events?} it is now:
\item
{\it How can we calculate the correlation including only photons of the same pair?}
\end{itemize}

The former question leads, and did in fact lead, to somewhat
ridiculous speculations about the metaphysical implications of
a physical theory. In our view something, the next generation
will laugh about. The latter question is only a technicality.
It can be solved by methods strictly within the limits of
theoretical physics. And it has already been solved by the
methods used in quantum mechanics. One only has to reinterpret
these methods as such: a technicality to guarantee that the
linearity of the fields between the polarizers is conserved.
Once this point is understood in quantum mechanics, a great deal,
maybe all, of the paradoxa will simply disappear.

In a sense, EPR experiments are similar to interference measurements:
because the result depends on the phase between the two locations.
Therefore, the main feature of the presented model, the individual
photon being in phase with the extended field, may be worth a
more general application. Then the wave-particle duality could
eventually be comprehended as an individual (the particle, the photon)
participating via its phase in some larger, possibly infinite structure
(the wave). What this would mean, historically, seems clear: a recovery
of Louis de Broglie's conception of a {\it harmony of phases}
on a different basis.

Incidentally, this expression for the correlation probability is
equal, but for a factor of 2, to the probability in the single
photon case. It seems that a time ordered representation of
measurements retains the linearity of fields, while it accounts
for correlations in a straightforward manner. The equivalence,
implied by semantics, has its counterpart in physics. The one
photon model is also the only case, where a single event can be
traced in a precise spacetime picture. A clear distinction between
single events and their statistical treatment, which is a feature
of this model, can contribute much to remove the paradoxa in
quantum mechanics. This has been already been emphasized by Barut
\cite{barut92}. Unfortunately, the lesson seems forgotten. And in
standard quantum mechanics, the distinction is not made in
principle.

Whether the imaginary part of the integral has a
significance of its own, e.g. related to the kinetic energy and thus
momenta of the photons (which are in principle measurable), cannot be
said with certainty at the moment. This point will be addressed in
future.

This result is in accordance with all experimental findings.
Physically speaking, the model is a local model of EPR-type
measurements, regardless of the actual separation of the two measuring
devices. And the only physical condition for the derivation are a relation
of the phase (or the angle of rotation) of the electromagnetic field of
photon 1 and photon 2, and a coherence length high enough to guarantee
the right phase relation between the finite fields of the photons and the
hypothetical field extending throughout the system.

Does the current model violate the Bell inequalities? It is easy to see, that it
does, because the joint probability
$\bar{P} (\alpha,\beta) = \frac{1}{2} \sin^2 (\alpha - \beta)$
contains the interference term.
Like in the conventional model truncating the phase information by a locality condition
yields disagreement between calculated and measured values. Although, in the present case,
the model is perfectly local. The main point, from a physical
perspective, is the difference between conceptual nonlocality - which exists, without
doubt, in quantum theory - and physical nonlocality - which requires a connection
explicitly contradicted by special relativity. Saying that, we also wish to make it clear,
that we do not enter into the more subtle points of the debate, how information is
transferred instantaneously from one device to the other without violating special
relativity.

Given the result of this calculation, we are faced by a considerable
problem. Bell states, in his paper, that \cite{bell64} '' in a theory,
in which parameters are added to quantum mechanics to determine the results
of individual measurements without changing the statistical predictions,
there must be a mechanism, whereby the setting of one measuring device can
influence the reading of another instrument, however remote.'' In this
model we found that the conditional probability of the measurement 2
given 1 necessarily involves the setting of 1, which expresses the
phase of the photon's electromagnetic fields. Our principal question
concerning the Bell inequalities can then be formulated as follows.
If (i) the result of our calculation is equal to the result of measurements,
and (ii) the calculation involves, necessarily, settings at two separate
locations, while (iii) the whole model is nevertheless strictly local, then:
{\it What is the logical flaw in Bell's acclaimed inequalities?}
Because the Bell inequalities, it should be remembered, are taken as a
proof that local and realistic models of EPR-like measurements do not
exist. Whereas this model, based on the phase of the electromagnetic fields,
is a local and realistic model.

Bell assumed the correlation function $P({\bf a}, {\bf b})$, to be given by
(in the general case the plane of polarization is described by a vector in space)
\cite{espagnat89}:

\begin{equation}
P ({\bf a}, {\bf b}) = \int d \lambda \rho (\lambda)
\bar{A} ({\bf a}, \lambda) \bar{B} ({\bf b}, \lambda)
\end{equation}

Here $\lambda$ is the hidden variable, {\bf a},{\bf b} the setting of
device 1, 2, and $\bar{A}$, $\bar{B}$ the expectation value of the
spin of photon 1 or 2. The only other assumption, used in the
derivation is the limitation of the expectation values $\bar{A}$, $\bar{B}$.
The main problem, with Bell's definition, is the separation of the two
photons. He assumed that the two systems can be described independently;
this is the conceptual meaning of the product
$ \bar{A} ({\bf a}, \lambda) \bar{B} ({\bf b}, \lambda)$.
The assumption is incorrect, considering the intrinsic electromagnetic
fields, if these fields are related by a phase. So that the
logical separation, Bell assumed, would only be given if we
measure two separate particles with random phases. In this
case the Bell inequalities are not violated.
The experiments with down-converted photons measure, in effect,
the phase relation between the two photons \cite{weihs98}.
While Bell, simply said, described a measurement with two independent
photons. That the phase relation lies at the bottom of the problem,
can also be demonstrated in a different, more abstract approach
\cite{unni00}.

One point seems to deserve special emphasis. Bell's statement is
frequently interpreted in terms of nonlocal connections between the
two {\it measuring devices}. Within the current model, this
interpretation is incorrect. Because the interaction between
a polarizer and a photon is always strictly local. It depends only on
two local variables (i) The setting of the polarizer, and (ii) the
orientation of the photon's electromagnetic field. Apart from the
phase due to the emission at a common source, there is no connection
between the measurement events. This can clearly be seen, if two
different outcomes are analyzed: correlations exist, or do not
exist between the two measurements. The only difference between
these two cases is the quality of the photons: in one case their
phase is related, in the other case it is not. Therefore, the
connection can be changed simply by changing the quality of the
photons. Nonlocality requires that the two events are connected
not only via the photons. Since the correlations change by changing
the photons, the claim of any nonlocal connection must be rejected.
The model is therefore local.

The current understanding of this important experiment can be used to shed new light
on the old questions, connected to Einstein's original paper \cite{epr35}.
Because the insight into the physics of the actual process {\it behind}
a measurement of polarization (or photon spin, in the terminology of quantum
mechanics), may be used to ask, whether Einstein was right, when he suspected
quantum theory to be incomplete \cite{epr35}? On the basis of this result,
we may conclude that he was. Because the measurement of the electromagnetic
field of a {\it single} photon and the phase relations between two photons is
well beyond a purely statistical interpretation of quantum theory. But so was
Bohr \cite{bohr36}: because there is
no defined quality of any single photon, which exists after the emission from a
common source. All that exists, is a periodic electromagnetic field in some
angle of polarization. The problem at their time seemed to be the assumption of
''objective'' properties of photons. And the rather simple concept of a particle.
A periodic field is no ''objective'' property, neither is the superposition of two
discrete states of polarization.

In summary, we have presented, what we consider the first step towards
a fully local model of EPR measurements. Locality was regained by an analysis of
the measurements within a field theoretical approach, and using a non standard
model of photons. It was essential, for the derivation, that the linearity
of electromagnetic fields between the polarizers was preserved.  The same
requirement, in quantum mechanics, goes under the name of {\it entanglement}.
It was shown that only a combined field theoretical and statistical
treatment provides sufficient constraints for mathematical models to
construct a representation of these measurements on
a local basis. We expect the same feature, correlations of polarization
measurements in space like separation, to arise in every
experiment, where joint probabilities of separate
measurements are determined.

\section*{Acknowledgements}

Helpful discussions with Guillaume Adenier and Al Kracklauer
are gratefully acknowledged.

%

%

\end{document}